\documentclass[%
 aip,
rsi,
amsmath,amssymb,
reprint,%
groupedaddress, % for one address of all authors
]{revtex4-1}

\usepackage{graphicx}% Include figure files
\usepackage{dcolumn}% Align table columns on decimal point
\usepackage{bm}% bold math
\usepackage[utf8]{inputenc}
\usepackage[T1]{fontenc}
\usepackage{mathptmx}

    \def\bank{p}
	\def\snub{\textrm{snub}}
	\def\sam{\textrm{sam}}

	\def\lead{\textrm{lead}}
    \def\bias{\textrm{bias}}
    \def\leak{\textrm{leak}}

    \def\peak{\textrm{peak}}

    \def\max{\textrm{max}}
    \def\leak{\textrm{leak}}    
    
    \def\term{\textrm{term}}

\begin{document}

%\preprint{Draft}
% \title{A kilowatt solid-state pulse amplifier for microsecond-timescale Joule heating experiments in diamond anvil cells}
\title{A broadband pulse amplifier for Joule heating experiments in diamond anvil cells}

\author{Zachary M. Geballe}
\author{Joseph Lai}
\author{Michael J. Walter}
\affiliation{%
Earth and Planets Laboratory, Carnegie Institution for Science, Washington, DC 20015 %\\This line break forced% with \\
}%

\date{\today}

\begin{abstract}
Decades of measurements of the thermophysical properties of hot metals show that pulsed Joule heating is an effective method to heat solid and liquid metals that are chemically reactive or difficult to contain. In order to extend such measurements to megabar pressures, pulsed heating methods must be integrated with diamond anvil cells. We report the design and characterization of a new pulse amplifier that can heat~$\sim 5$~to~$50$~$\mu$m-diameter metal wires to 1000s of kelvin at megabar pressures using diamond anvil cells. Pulse durations and peak currents can each be varied over 3 orders of magnitude, from 5~$\mu$s to 10 ms and 0.2 to 200 A. The pulse amplifier is integrated with a current probe. Two voltage probes attached to the body of a diamond anvil cell can be used to measure voltage in a four point probe geometry. The accuracy of four point probe resistance measurements for a dummy sample with 0.1~$\Omega$~resistance are typically better than~$5 \%$~at all times from 2~$\mu$s to 10 ms after the beginning of the pulse. 
\end{abstract}

\maketitle

\section{Introduction}
Continuous laser heating, pulsed laser heating, and continuous Joule heating in diamond anvil cells have been used for a wide variety of melting,\cite{Weir2012,Sinmyo2019,Parisiades2021}, synthesis,\cite{Alabdulkarim2023} and transport measurements\cite{Yagi2011,Konopkova2016} up to several megabars of pressure. Until very recently, pulsed Joule heating had not been combined with DACs, despite the success of pulsed laser methods,\cite{Yagi2011,Konopkova2016,Zaghoo2016} and despite the finite element model results showing it might provide an unambiguous way to detect melting transitions.\cite{Geballe2012}

In contrast, pulsed Joule heating has been used for decades of groundbreaking measurements on the thermophysical properties of materials heated from ambient pressure to temperatures far above melting.\cite{Cagran2008,Cezairliyan1987,Kondratyev2019} Typically, a pulse of current with~$\sim$kA~amplitude is used to heat a mm-diameter sample to many 1000s of K on the~$\mu$s to ms timescale. \cite{Gallob1986,Cagran2008,Cezairliyan1987,Henry1972} Historically, pulse amplifiers were based on vacuum tube technologies such as triggered spark gaps or ignitrons because of their high current and voltage ratings ($>$ kA, kV).\cite{Cook2000} Typically, two spark gaps were used: one to start the pulse and a second one to end the pulse. Two are required because spark gaps are ``semi-controlled'' switches, meaning they can only be externally controlled in one direction, on or off. Moreover, each spark gap required a high voltage ($\sim$kV) trigger, which was typically provided by yet another vacuum tube (e.g. a thyratron), adding complexity to the circuit.\cite{Henry1972,Gallob1986}

Recently, we showed that pulses with smaller pulse amplitude (10s of A, 10s of V) can be used to melt platinum samples on the~$\mu$s-timescale at pressures up to 1.1 megabar in diamond anvil cells. \cite{Geballe2021} Because of the relatively low power requirements, we were able to use a simple pulse amplifier circuit based on a ``fully-controlled'' solid-state switch that was triggered directly from a delay generator. Moreover, the pulse amplifier was small, safe, and robust enough to transport to a synchrotron facility for X-ray measurements synchronized with the Joule heating pulses.

However, three limitations of the pulse amplifier and sensing circuitry were detected during the platinum heating experiments.\cite{Geballe2021} First, the pulse power was limited by the 60 V maximum rating of the N-channel MOSFET around which the amplifier was based. Second, at least one of the instrumentation amplifiers used for current and voltage sensing seemed to become damaged, causing spurious distortions in our final set of measurements (Section S1c of Ref. \onlinecite{Geballe2021}). Third, the accuracy of the voltage drop across the sample was mediocre in cases where the sample resistance was much smaller than the series resistance of the MOSFET.

Here, we describe a new pulse amplifier and electrical sensing equipment designed for improved robustness, voltage range, and accuracy of current and voltage measurements during pulsed Joule heating. The crucial new components are a high-side solid state switch and passive probes, rather than the N-channel MOSFET and instrumentation amplifiers used in Ref. \onlinecite{Geballe2021}.

\begin{figure*}[tbhp]
	\centering
	\includegraphics[width=6in]{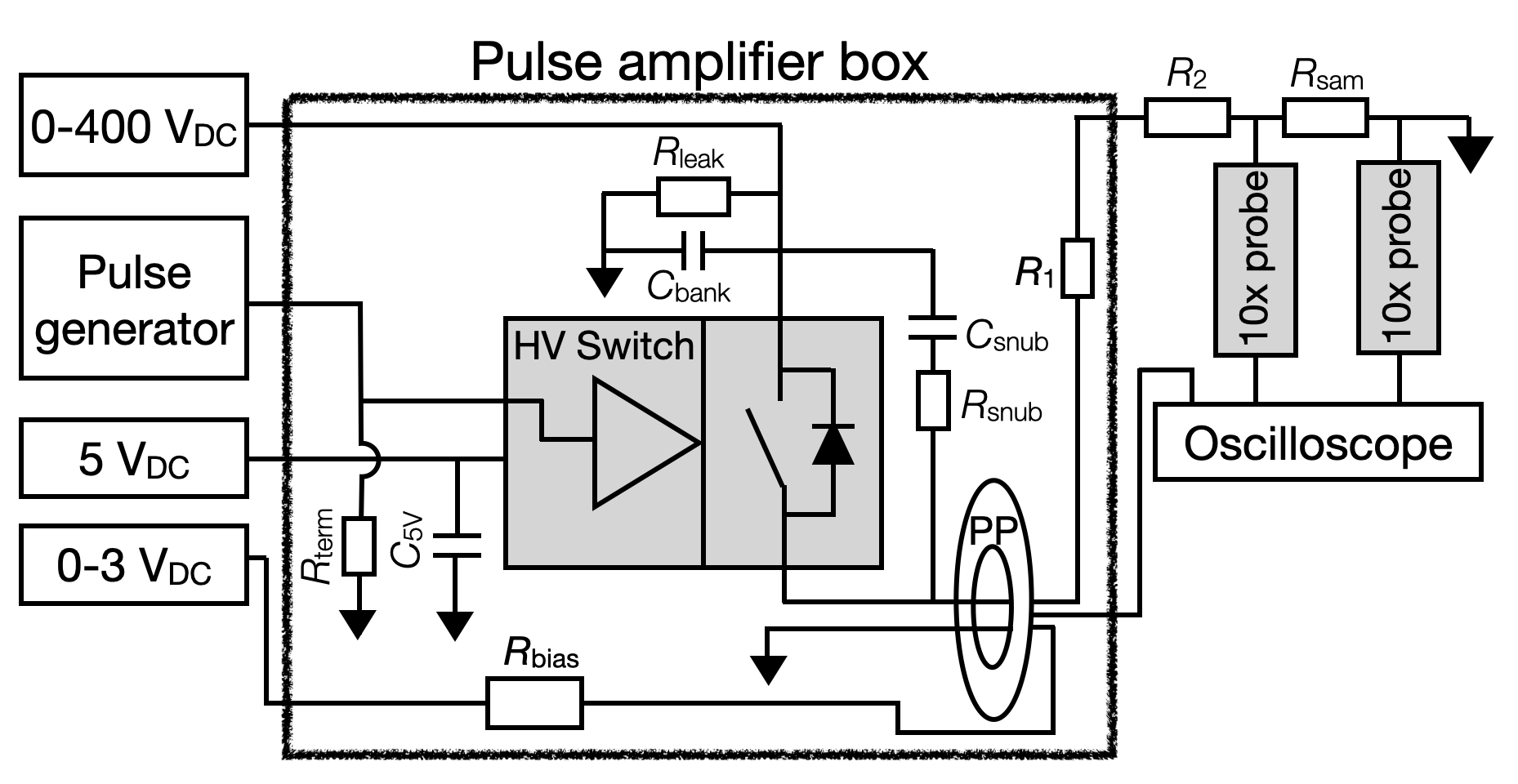} %5.9 in 
	\caption{Schematic of the pulse amplifier, current probe, and voltage probes. The large black box contains the electrical elements that are placed inside the metal enclosure.  The shaded rectangles inside the pulse amplifier box shows the schematic of the Behkle high voltage switch. Voltage and current probes are marked by the shaded boxes labeled ``10x probe'' and the donut labeled ``PP''. Part numbers are listed in Table \ref{table:parts}.}
	\label{fig:schematic}
\end{figure*}

\section{Electrical Design}
\begin{figure*}[tbhp]
    \centering
    \includegraphics[width=6.3in]{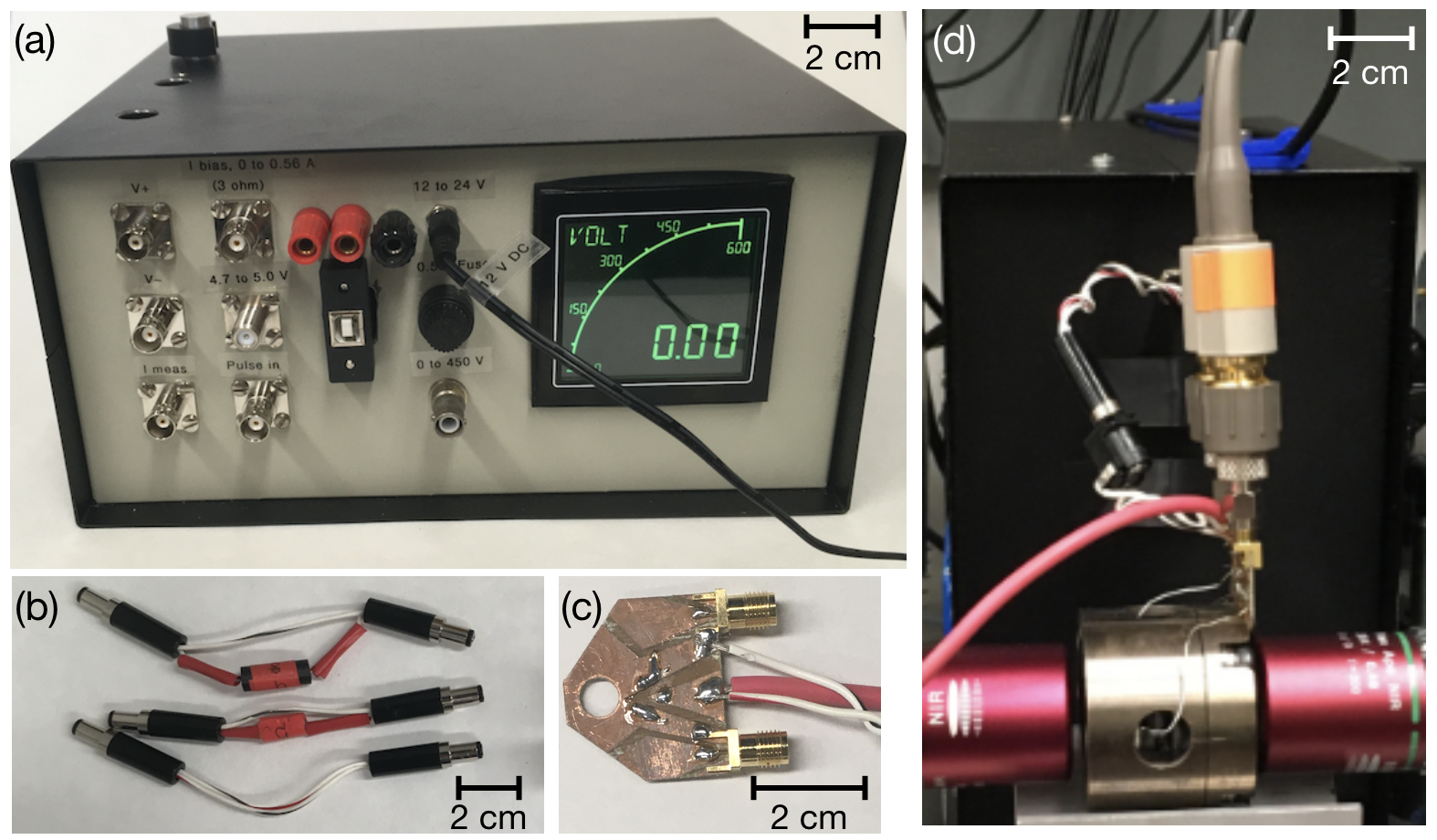} %5.9 in 
    \caption{Photos of electrical equipment and high pressure equipment. (a) The front of the pulse amplifier. Several jacks were used for testing and are not used in this study. (b) Electrical leads that connect the pulse amplifier box to the sample. Two leads include  resistors with~$R_\lead = 10$ or 3~$\Omega$. (c) The small circuit board that is used to connect the diamond cell to the electronics. (d) The pulse amplifier and diamond cell in action. The pulse amplifier rests sideways on an optical table. The diamond cell is held on a V-block between objective lenses. The small circuit board is screwed on to the diamond cell, and connected to the high pressure sample by thin wires that enter through the diamond cell portholes. Two 10x oscilloscope probes are connected to the small circuit board by SMA connectors.}
    \label{fig:photos}
\end{figure*}

The pulse amplifier is designed around the HTS-81-25-B switch from Behlke (Fig. \ref{fig:schematic}). The switch is powered by 5 V and 0.2 A from a regulated DC power supply. A DC power supply charges the primary capacitor, ``$C_\bank$'', which stores charge at the high voltage side of the switch. When the switch is triggered by a TTL signal (2-5 V amplitude) from a pulse generator, the capacitor is discharged through a series of buffer resistors and through the sample to ground. The first buffer resistor,~$R_1$, is a 1~$\Omega$,~25 W resistor that is hard-wired inside the pulse amplifier box.  The second buffer resistor,~$R_2$, is twisted into the delivery leads (Fig. \ref{fig:photos}b), allowing the user to easily switch between values of~$R_2$. The data presented here uses~$R_2 = 0$~or 3~$\Omega$. The discharge is cut off when the TTL signal ends. The natural decay timescale of the capacitor discharge is~$\tau = RC = (R_1 +R_2 + R_\sam) \times 70$~$\mu$F, which equals 70~$\mu$s for~$R_2 = 0$~$\Omega$~and 280~$\mu$s for~$R_2 = 3$~$\Omega$. In other words, the pulse current decays exponentially to the maximum current of the DC power supply in 100s of~$\mu$s.

Electrical sensing relies on three passive probes: a broadband current monitor (also known as a ``Pearson probe''; labeled ``PP''), and two 10x oscilloscope probes. Part numbers are listed in \ref{table:parts}. The Pearson probe outputs a voltage to the oscilloscope that is equal to 0.1 times the current. Hence, the oscilloscope measures~$0.1 I$,~$0.1  V_+$, and~$0.1 V_-$, where~$V_+$~and~$V_-$~are four point probe voltages at the high side and low side of sample, respectively. (After collecting the data for this manuscript, a third voltage probe was added to monitor voltage at the output of the pulse amplifier box, which is convenient for monitoring resistance in a two point probe configuration.) 

Note that the use of a high-side switch enables the use of a passive voltage probe to measure accurate values of~$\Delta V = V_+ - V_-$. If we had used a low-side switch and passive voltage probes, each probe would measure the voltage of interest \textit{plus} the voltage drop across the switch, which is often much larger than the voltage of interest. For example, if the switch's series resistance is 1~$\Omega$~and the high pressure metal sample's resistance is~$\frac{1}{50}$~$\Omega = 20$~m$\Omega$, a 1\% error in measurement of $V_+$ or $V_-$ propagates to $\sim 50$\% error in~$\Delta V$.

Two additional components help avoid saturation of the electrical measurements at high current and voltage. First, if the integral of current over time exceeds~$\sim 0.02$~Amp-seconds for a single pulse, Pearson probe Model 411 saturates according to the manufacturer's specification sheet. To extend the time-current integral for the highest current pulse presented here, we follow the manufacturer's recommendation to run 0.1 A of continuous direct current in the reverse direction through a very simple circuit that consists of a DC power supply and a 3~$\Omega$~resistor labeled ``$R_\bias$''. Second, if the Pearson probe's output voltage exceeds the 50 V range of the oscilloscope, we add a 10 dB attenuator at the oscilloscope input. 

Additional resistors and capacitors are added to the pulse amplifier to damp oscillations. A 47~$\Omega$~resistor labeled ``$R_\term$'' in Fig. \ref{fig:schematic} terminates the output from the pulse generator. A 10~$\mu$F capacitor buffers the 5 V DC power supply. A capacitor and resistor in series create a low-pass filter, which damps high frequency oscillations that would otherwise oscillate across the switch when it is turned on and off. They are labeled ``$C_\snub$'' and ``$R_\snub$''. Their product,~$R_\snub C_\snub = 1 \Omega \times 50$~$\mu$F~$=50$~ns is the low pass filter timescale. 

Finally, two features are added for safety purposes. First, a 250 k$\Omega$~resistor, ``$R_\leak$'', drains the capacitor ``$C_\bank$'' in the characteristic time~$R_\leak C_\bank = 17$~seconds. Second, a panel-mount voltmeter monitors the voltage on the high side of the capacitor. The voltmeter is shown in the photo in Fig. \ref{fig:photos}a, but not in the schematic.

\section{Mechanical Design and Construction} 
A 25 x 20 x 13 cm metal box houses a circuit board that is held on four standoffs. Five elements are attached by screws to the circuit board: the Pearson probe, the HV switch, and three power resistors ($R_\leak$,~$R_\bias$~and~$R_1$).The primary capacitor,~$C_p$,~is pressed into through holes and soldered. The other elements are attached by solder. The faces of the box hold the panel-mount voltmeter, panel-mount BNC connectors for connections with the DC power supplies, pulse generator, and oscilloscope, and a panel-mount barrel connector that connects to~$R_\lead$~and~$R_\sam$ to the pulse amplifier. The oscilloscope probes are fastened onto a face of the box for strain relief.

\begin{figure*}[tbhp]
	\centering
	\includegraphics[width=6.3in]{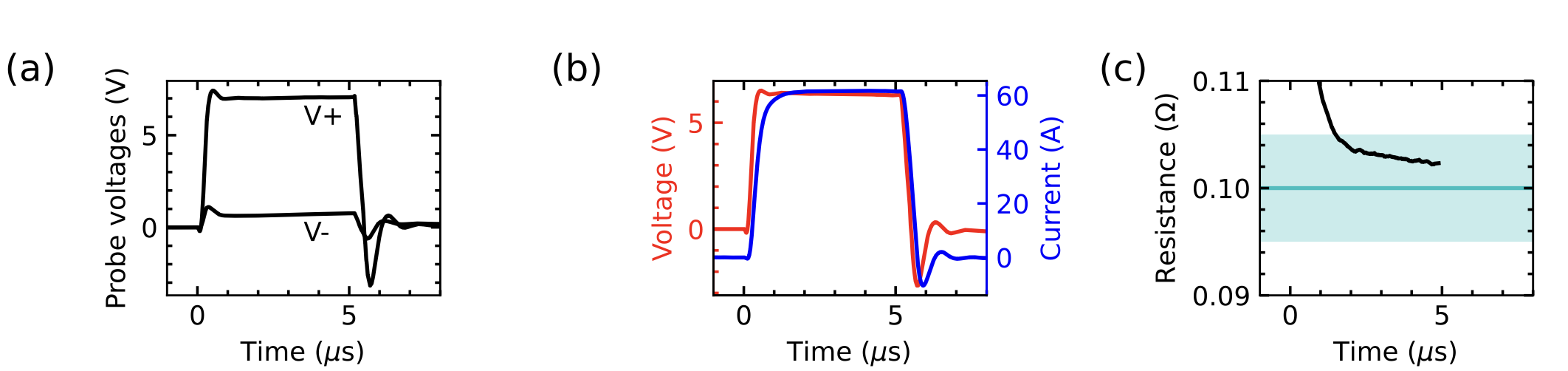} %5.9 in 
	\caption{A 5~$\mu$s pulse to 60 A through the 100 m$\Omega$~dummy sample. (a) Voltage measurements from electrodes that probe just above and just below the dummy sample in a four point configuration. (b) Current measured by the Pearson probe (blue) and four point probe voltage (red) calculated by the difference between curves in (a). (c) Pulsed resistance measurement (black) and actual 100 m$\Omega$~resistance (cyan line)  with~$\pm 5$~\%error envelope (cyan shading). Settings:~$R_2= 3$~$\Omega$, SRS power supply, 0 A through~$R_\bias$.}
	\label{fig:one_example}
\end{figure*}

\section{Results}
\subsection{100 m$\Omega$~dummy sample}
The pulse amplifier is tested with~$R_\sam = 100$~m$\Omega$~and~$R_2 = 0$~or 3~$\Omega$. The 100 m$\Omega$~dummy sample is Ohmite brand, part number WHER10FET. It has a 5 W power rating and a nominal precision of 1\% in resistance, which we confirm by measurement with a Keithley Sourcemeter 2400 before and after pulsing high amplitude currents through it. Single pulses are used during 100 m$\Omega$~dummy testing (i.e. no averaging to reduce noise). The start and end of each pulse is triggered by a pulse generator (Berkeley Nucleonics Model 525).

An example of a 5~$\mu$s pulse using~$R_2 = 3$~$\Omega$~is shown in Fig. \ref{fig:one_example}. The current rises to~$I_\max = 60$~A with a rise time of 460~$\mu$s, and drops to 0 A with a fall time of 350~$\mu$s. Rise time and fall time are defined by time interval between 10\%~$I_\max$~and 90\%~$I_\max$. Resistance is determined by dividing the four point probe voltage by current. In this example, the accuracy of resistance is better than  5\%  at all times from 1.4 to 4.9~$\mu$s.

\begin{figure*}[tbhp]
	\centering
	\includegraphics[width=4in]{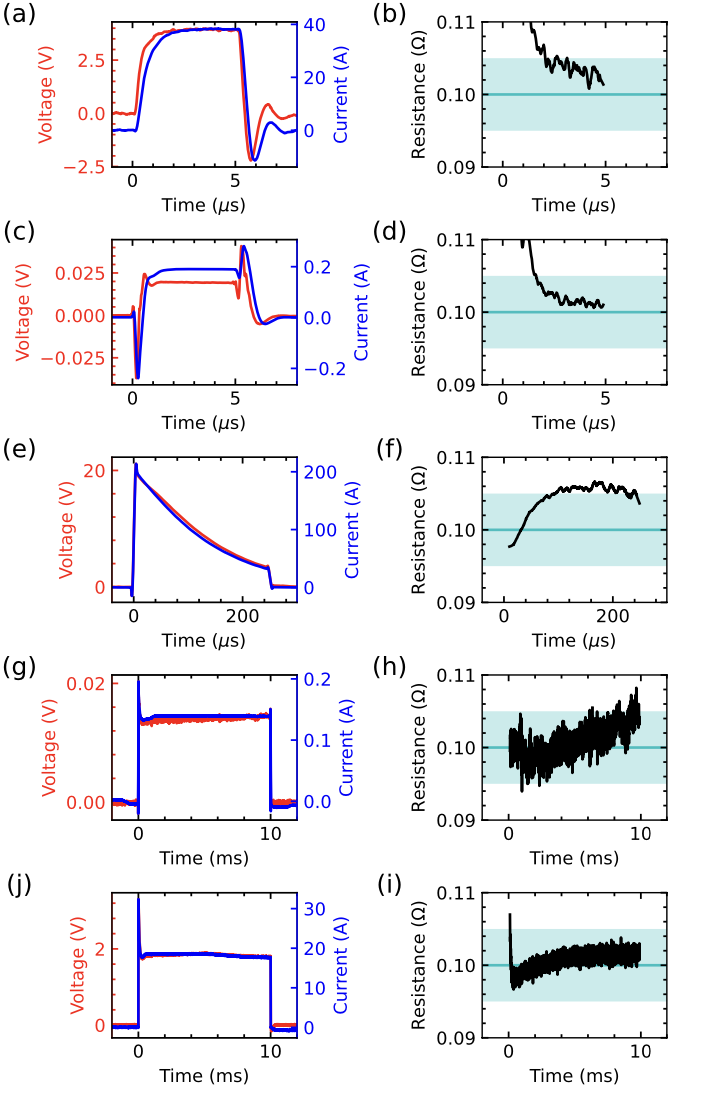} %5.9 in 
	\caption{Five pulses through the 100 m$\Omega$~dummy sample. Each row shows the four point probe voltage (red), current (blue), and resistance (black). (a,b) A 5~$\mu$s pulse to 35 A using~$R_2 = 0$~$\Omega$~and the SRS power supply. (c,d) A 5~$\mu$s pulse to 0.2 A using~$R_2 = 3$~$\Omega$~and the SRS power supply. (e,f) A 250~$\mu$s pulse peaking at 200 A using~$R_2 = 0$~$\Omega$, the SRS power supply at 400 V, and 0.1 A DC current through~$R_\bias$. (g,h) A 10 ms pulse to 0.14 A using the~$R_2 = 3$~$\Omega$~and the XHR power supply. (j,i) A 10 ms pulse to 18 A using~$R_2 = 0$~$\Omega$~and the XHR power supply.}
	\label{fig:extreme_examples}
\end{figure*}

Fig. \ref{fig:extreme_examples} shows four other tests that demonstrate the range of pulses that can be driven from the pulse amplifier. The pulse amplifier can operate over 3 orders of magnitude in pulse duration and in peak current: 5~$\mu$s to 10 ms, and 0.2 to 200 A. In almost all cases, resistance measurements are within 5\% of the true value at all times from~$\sim 2$~$\mu$s until the end of the pulse (Fig. \ref{fig:extreme_examples}b,f,h). The main exception is that the error reaches~$\sim 6\%$~during the highest amplitude pulse (Fig. \ref{fig:extreme_examples}d). Note that to achieve the longest pulses, we used a high current DC power supply (XANTREX XHR 60-18), whereas to achieve relatively short, high voltage pulses we used an SRS PS310 power supply to charge~$C_\bank$. Also note that if we attempt to end the pulse when current exceeds~$\sim 60$~A, the circuit begins to oscillate wildly; high amplitude 500 kHz oscillations persist for more than 100~$\mu$s. These oscillations ringing could damage the pulse amplifier or a delicate sample. Hence, the amplitude of the pulse in Fig. \ref{fig:one_example} cannot be increased without risk, and the duration of the pulse in Fig.  \ref{fig:extreme_examples}c cannot be \textit{decreased} by more than~$\sim 100$~$\mu$s without risk. 

The longest duration pulse that can be measured with high accuracy in our measurement scheme is 10 ms. At longer times, Pearson probe Model 411 approaches its low frequency bandwidth limit. The manufacturer's specification sheet lists a 0.9 \%/ms ``droop rate'', and indeed at 10s of ms, the measured current decreases steadily over time by more than 10\% even though the voltage measurement remains steady. Numerical correction for the droop rate might be possible if a calibration of droop rate were performed.

Using~$R_2 = 0$~$\Omega$, the rise time is 1~$\mu$s. Using~$R_2 = 3$~$\Omega$, a faster rise time of~$\sim 500$~ns is achieved (Figs. \ref{fig:extreme_examples}a,c). Further increasing~$R_2$~to 10 or~$25$~$\Omega$ has no effect on the rise time. The rise time is also sensitive the snubber components ($C_\snub$,~$R_\snub$). By choosing a smaller values of~$C_\snub$~or~$R_\snub$, the rise time increases by a small amount, but the stability of the pulse amplifier is reduced. For example, decreasing~$C_\snub$~to 25 nF reduces the rise time to~$\sim 350$~ns, but the rising edge of the pulse includes a few oscillations. For~$C_\snub = 10$~nF, the oscillation amplitude increases further for both the rising edge and the falling edge, and the sharpness of the rising edge is essentially unchanged. 

\begin{figure}[tbhp]
	\centering
	\includegraphics[width=2in]{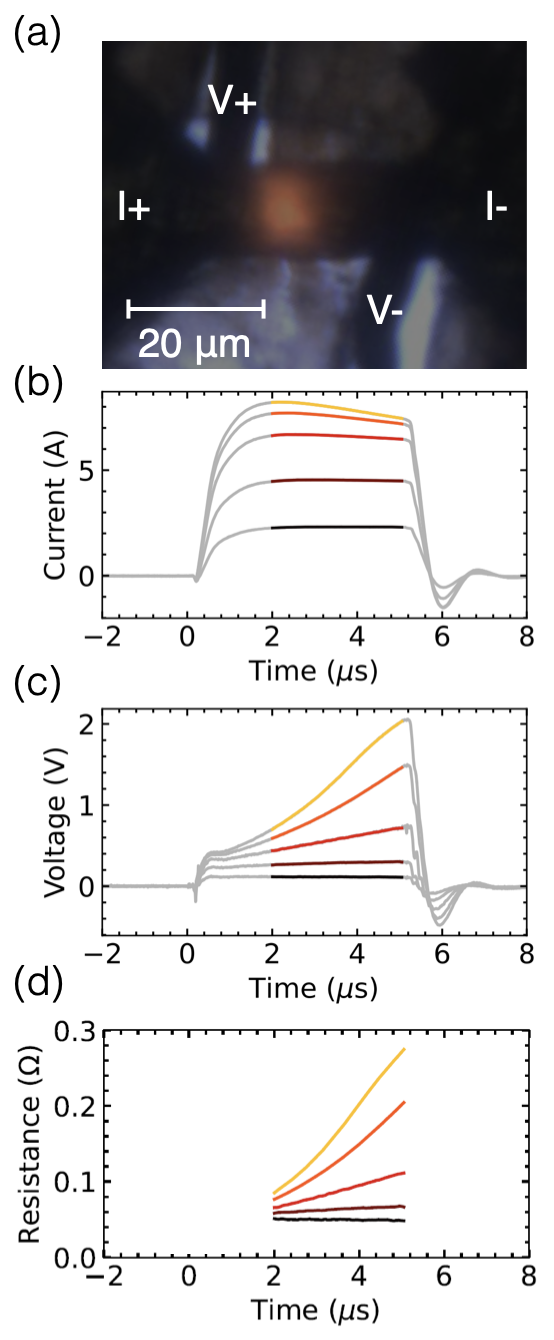} %5.9 in 
	\caption{(a) Two photographs overlayed: a white light image of a piece of iron compressed to 77 GPa in a diamond anvil cell, and a image of the same piece of iron being heated by 500 repetitions of 5~$\mu$s-long electrical pulses. The orange glow is the thermal emissions from the heated sample. ``V+'' and ``V-'' mark the voltage probes. ``I+'' and ``I-'' mark the electrodes that deliver current. (b) Current during five sets of 500 pulses, each of which uses a driving voltage from 5 to 19 V. (b) Voltage during the same sets of pulses. (d) Resistance during the same sets of pulses. Colors in (b) and (c) mark the data used to infer resistance in (d).}
	\label{fig:TheMistake}
\end{figure}

\subsection{Iron sample at high pressure}
We demonstrate the utility of the pulse amplifier by using it to heat a sample of iron compressed statically to 77 GPa at room temperature. The iron sample is shaped into a four point probe geometry and compressed between layers of KCl insulation in the sample chamber of a diamond anvil cell. The sample preparation method is described in Ref. \onlinecite{Geballe2023Microassembly}. 

We drive 5~$\mu$s pulses of current through the sample. At pulse amplitudes~$> 5$~A, the sample visibly heats to~$\sim 2000$~K, as evidenced by the orange glow in Fig. \ref{fig:TheMistake}a, and by spectroradiometry temperature measurements described in Ref. \onlinecite{Geballe2023PMT}. To generate clear photos of the glow, we accumulate light from 500 heating pulses at a 1000 Hz repetition rate. A composite image of the hot sample and the sample photographed in white light shows that the sample glows orange in part of the region between voltage probes (Fig. \ref{fig:TheMistake}a).

Examples of pulsed heating are shown in Figs \ref{fig:TheMistake}b-d. At low current amplitude, the resistance is steady. At higher current amplitudes, the resistance increases more than 3-fold during the pulse as the sample heats to 1000s of K.

\begin{figure}[tbhp]
	\centering
	\includegraphics[width=3.5in]{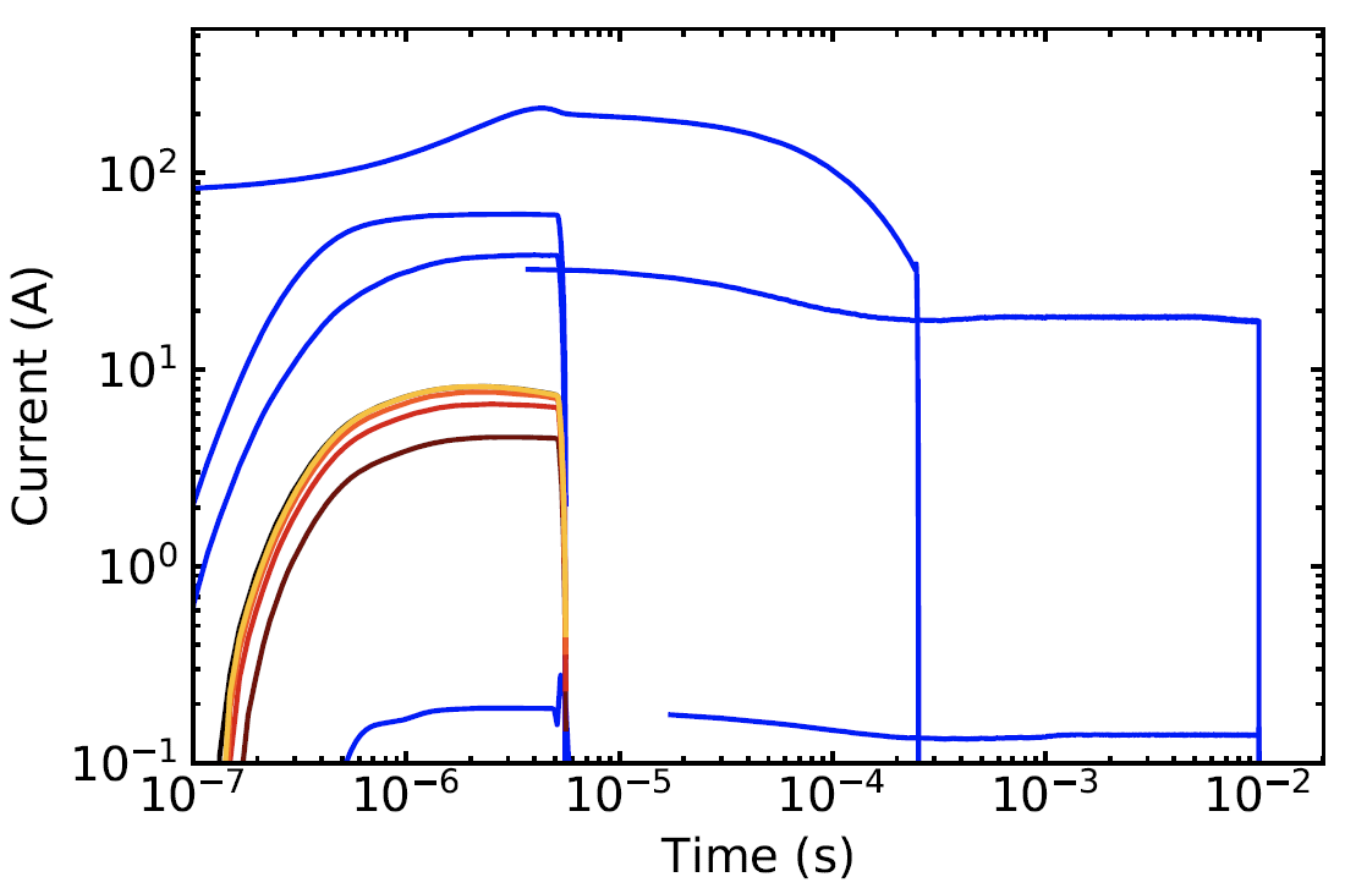} %5.9 in 
    \caption{Current versus time with log scaling for all data shown in Figs. \ref{fig:one_example}, \ref{fig:extreme_examples}, \ref{fig:TheMistake}. Blue curves show the current pulses through the 100 m$\Omega$~dummy sample. Black, brown, red, orange, and yellow curves show the current pulses through the high pressure iron sample. Data is truncated for clarity. For example, the long duration pulses are truncated at times shorter than the time-resolution of the data.}
	\label{fig:loglog_examples}
\end{figure}

\section{Discussion}
To summarize the wide range of pulses described above, we overlay many pulses in a log-log plot (Fig. \ref{fig:loglog_examples}).

To give ideas of how the pulse amplifier can be used, we describe pulses in relationship to a ``prototypical pulse'' for diamond cell experiments, which we define as the pulse plotted in yellow in Fig. \ref{fig:loglog_examples}. It is a 5~$\mu$s pulse that peaks at~$\sim 8$~A and causes the 77 GPa iron sample glow orange. The pulse amplitude can be decreased 40-fold to 0.2 A, a change that can be useful to avoid heating a sample while measuring its resistance. The pulse amplitude can be increased 7-fold to 60 A, or lengthened 200-fold to 10 ms, changes which are crucial for heating thicker or wider samples.

If a decaying pulse shape is acceptable, the prototypical 8 A pulse can be increased 25-fold to 200 A. This pulse is useful to melt relatively-thick samples at low pressure. For example, we used a pulse with~$I_\peak = 70$~A and 100$\mu$s duration to melt an ambient pressure iridium sample.\cite{Geballe2023PMT}

Shorter duration pulses are possible by triggering the switch with TTL signals that are shorter than 5~$\mu$s. As duration decreases, several issues arise. First, at~$\sim 2$~$\mu$s pulse duration, the plot of current-versus-time is rounded, not square, because the current rise and fall begin to overlap, at least for the~$\sim 100$~m$\Omega$~dummy sample. At 1~$\mu$s pulse duration, the error in resistance measurement becomes huge ($> 100$\%). The voltage probes measure spurious signals during the rise and fall pulse, likely caused by inductance of the sample and current-carrying electrodes \cite{Kaschnitz1992}. To correct for the inductance spikes, the method proposed in Ref. \onlinecite{Kaschnitz1992} could be useful. For trigger durations shorter than 1~$\mu$s, the pulse amplifier fails to turn on completely, because the rise time and fall time of the drive circuitry is too slow. We attempted to decrease the rise and fall times substantially by decreasing~$R_\snub$~and/or~$C_\snub$~by 2-fold and 5-fold, but with no success. The rise time remained~$> 300$~ns. In addition, the rising edge begins to oscillate at low~$R_\snub C_\snub$. The~$\sim 300$~ns lower limit for the rise time is surprising, because the Behlke switch has a nominal rise time of~$\sim 10$~ns. Albeit, the testing reported on the manufacturer's specification sheet is at much higher current and voltage than we use -- a minimum of 25 A and 840 V.

\section{Conclusions}
A broadband pulse amplifier has been built and characterized. It can heat metals in diamond anvil cells up to at least 77 GPa and 2000 K. If combined with well-prepared samples, heating to 1000s of K may be possible at several megabars of pressure. The pulse amplifier delivers 0.2 to 200 A with pulse durations in the range 5~$\mu$s - 10 ms.  When coupled with passive current and voltage probes, the accuracy of the typical four point probe resistance measurement is better than 5\%. The new pulse amplifier provides a more reliable and more versatile substitute for the amplifier used previously for pulsed Joule heating experiments up to megabar pressures.\cite{Geballe2021} 

\begin{acknowledgments}
This material is based upon work supported by the National Science Foundation under Grant No. 2125954. We thank Seth Wagner for machining parts. We thank Maddury Somayazulu for fruitful discussions.
\end{acknowledgments}

\section*{Data Availability Statement}
The data that support the findings of this study are available from the corresponding author upon reasonable request.

\begin{table}
\caption{List of parts, important specifications, and approximate price. Not included: metal casing, connectors, cables, circuit board, panel-mount voltmeter.}
\footnotesize{}
\begin{ruledtabular}

 % From Pulser_tables spreadsheet in LaTex_Files/
\begin{tabular}{	c		c		c		c		c	c		c	}
													
	Label	&		Specifications &  	&	Brand and	part number	\\
\hline	Oscilloscope	&	200 MHz	&	4 channels		& 	Picoscope	5444D	\\
	Pulse Generator	&	50 MHz	&	TTL	& 	BNC 525		\\
	HV supply (option 1)	&	1250 V	&	20 mA	& 	SRS	PS310 \\
	HV supply (option 2)	&	60 V	&	18 A &	XANTREX XHR 60-18		\\
    HV Switch   &   250 A$_\peak$   &   8.4 kV      &   Behlke  HTS-81-25-B  \\
    PP  &   20 MHz  &   5000 A$_\peak$  &   Pearson 411 \\
	DC supply	&	30 V	&	5 A		& 	Madell	TPR3003\\
    10x Probe & 400 MHz & 4 meters & Probemaster 6143-4 \\
$   C_\bank$    &   70~$\mu$F   &   500 V       &   TDK B32758G2706K000 \\
$   C_\snub$    &   50 nF   &   500 V   &       \\
$   R_\bias$    &   3~$\Omega$  &   50 W    &   Ohmite  850F3R0E      \\
$   R_1$    &   1~$\Omega$  &   10 W    &   Ohmite  810F1R0E      \\
$   R_\leak$    &   250 k$\Omega$   &   50 W    &   Ohmite  L50J250KE     \\
$   R_\sam$ (dummy) &   0.1~$\Omega$    &   5W  &   Ohmite  WHER10FET  \\
$	C_\textrm{5V}$	&	10~$\mu$F	&	25 V	& 	Nichicon	UPS1E100MDD	\\

$	R_{\textrm{term}}$	&	47~$\Omega$	&	1/4 W	& 	Ohmite	OD470JE \\
$	R_\snub$	&	1~$\Omega$	&	1/2 W	& 			\\

\end{tabular}

\end{ruledtabular}
\label{table:parts}

\end{table}

\clearpage
%\nocite{*}
\bibliography{Pulser}% Produces the bibliography via BibTeX.

\end{document}